\documentclass[twocolumn,english,prl, showpacs, superscriptaddress]{revtex4-1}
\usepackage[T1]{fontenc}
\usepackage{color}
\usepackage{amsmath}
\usepackage{amssymb}
\usepackage{graphicx}
\usepackage{wasysym}

\makeatletter

\providecommand{\tabularnewline}{\\}

\usepackage{hyperref}
\hypersetup{
    colorlinks=true,
    linkcolor=red,
    citecolor=blue,
    filecolor=blue,      
    urlcolor=blue,
}
\makeatother

\usepackage{babel}
\begin{document}
\title{\noindent \textcolor{black}{Universal two-level quantum Otto machine
under a squeezed reservoir }}
\author{Rogério J. de Assis}
\address{Instituto de Física, Universidade Federal de Goiás, 74.001-970, Goiânia
- GO, Brazil}
\author{José S. Sales}
\address{Campus Central, Universidade Estadual de Goiás, 75132-903, Anápolis,
Goiás, Brazil}
\author{Jefferson A. R. da Cunha}
\address{Instituto de Física, Universidade Federal de Goiás, 74.001-970, Goiânia
- GO, Brazil}
\author{Norton G. de Almeida}
\address{Instituto de Física, Universidade Federal de Goiás, 74.001-970, Goiânia
- GO, Brazil}
\begin{abstract}
We study an Otto heat machine whose working substance is a single
two-level system interacting with a cold thermal reservoir and with
a squeezed hot thermal reservoir. By adjusting the squeezing or the
adiabaticity parameter (the probability of transition) we show that
our two-level system can function as a universal heat machine, either
producing net work by consuming heat or consuming work that is used
to cool or heat environments. Using our model we study the performance
of these machine in the finite-time regime of the isentropic strokes,
which is a regime that contributes to make them useful from a practical
point of view.
\end{abstract}
\maketitle

\section{Introduction}

\begin{figure}
\centering{}%
\begin{tabular}{cc}
\includegraphics[width=8cm]{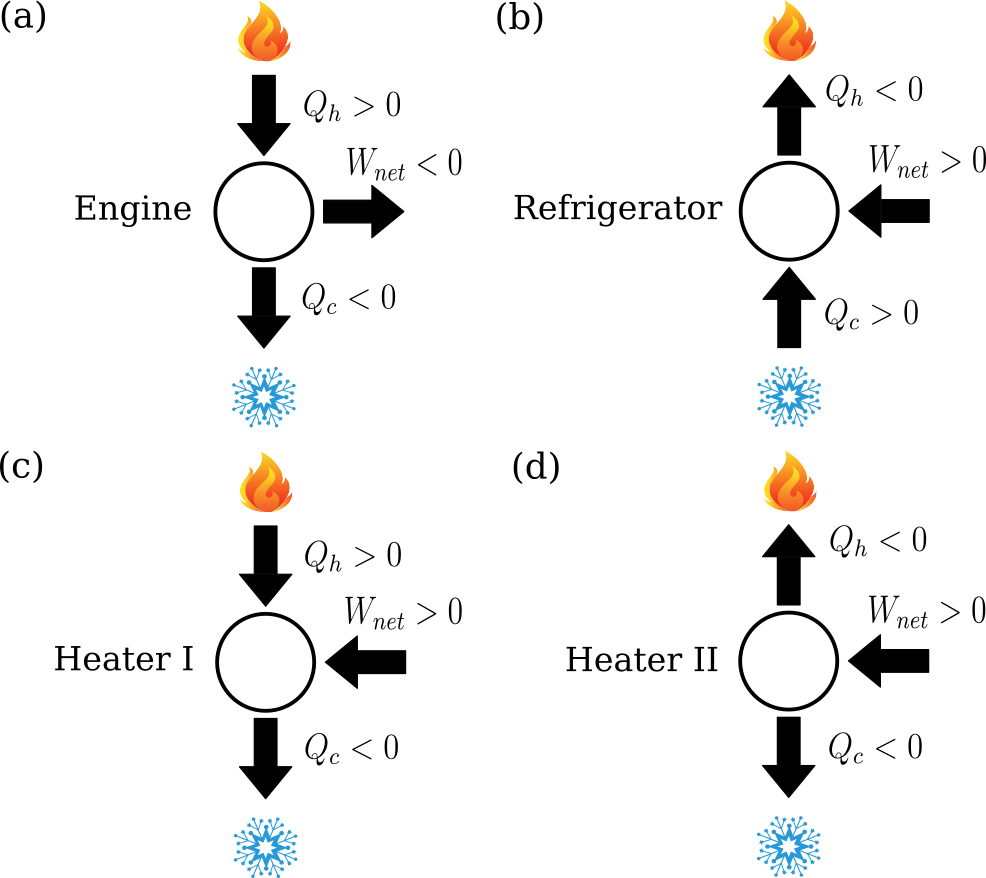} & \tabularnewline
\end{tabular}\caption{\label{Fig1} The four types of heat machines: (a) a heat engine,
which draws heat from the hot reservoir and dumps it in the cold one,
performing work, (b) a refrigerator, on which work is performed and
used to draw heat from the cold reservoir and pump it to the hot one,
(c) a heater that uses the work done to pump heat from the hot to
the cold reservoir, and (d) a second type of heat machine that uses
the work done to heat two environments at once.}
\end{figure}

Classical heat machines convert thermal resources into work and \emph{vice-versa}.
As for example, the heat engine draws heat from a hot reservoir, uses
part of that heat to perform mechanical work and discards the rest
in a cold reservoir. The refrigerator, on the other hand, uses mechanical
work to remove heat from a cold reservoir and discard it in a hot
one. A third kind of heat machine, the heather, uses the mechanical
work to heat one, usually the cold, or both reservoirs - see Fig.
\ref{Fig1}. The cyclic heat machines are the paradigm for these comparative
studies, whose efficiency $\eta$ for heat engines and coefficient
of performances $COP$ for refrigerators are related by
\begin{equation}
COP=\frac{1}{\eta}-1.\label{eq:eq1}
\end{equation}
Eq. (\ref{eq:eq1}) answers the following question: given a cyclic
heat machine operating reversibly, if a work $W$ is extracted with
efficiency $\eta$, what is the performance coefficient $COP$ if
that same heat machine operates in a reverse cycle consuming a work
$-W$. The heat machine efficiency and performance coefficient are
bounded by the Carnot efficiency and performance coefficient, which,
for thermal reservoirs, are only attained in quasi-static or reversible
cycles. It is currently a subject of intense study to compare heat
machines running on purely classic resources with heat machines running
on some kind of genuinely quantum resource, as for example coherence
\citep{Scully2003,Scully2011,Rahav2012,Uzdin2015,Turkpen2016,Brandner2017,Dorfman2018,Camati2019},
entanglement \citep{Wang2009,Correa2013,Brunner2014,Perarnau-Llobet2015,Tacchino2018},
as well as exploring the finite dimension of Hilbert space \citep{Kosloff2000,Linden2010,Gelbwaser-Klimovsky2013}. 

A heat machine whose working substance is a quantum system is often
called a quantum heat machine. Potential technological applications
of quantum heat machines ranges from heat transport in nano devices
\citep{Bermudez2013,Ronzani2018} to biological process control \citep{Muller1983,Mallouk2009},
among others \citep{Johnson2014}. One kind of quantum heat machine
widely addressed by researchers in the field of quantum thermodynamics
is the quantum Otto heat machine (QOHM) \citep{Linden2010,Wang2012,Vinjanampathy2016,Karimi2016,Abah2016,Kosloff2017,Peterson2019,Lee2020}.
\textcolor{black}{The QOHM }consists of two isochoric strokes, one
with the working substance coupled to the cold thermal reservoir and
the other coupled to the hot thermal reservoir, and two isentropic
strokes, in which the working substance is disconnected from the thermal
reservoirs and evolves unitarily\textcolor{black}{. }In the past years,
non-thermal reservoirs have also been used in the theoretical and
experimental study of QOHM, \textcolor{black}{for instance squeezed
thermal reservoirs \citep{RoBnagel2014,Long2015,Manzano2016,Klaers2017,Assis2020,Singh2020}
and reservoirs at apparent negative temperature \citep{Assis2019}.
These unconventional quantum engines have drawn attention due to the
promising gains in engine efficiency and power.}

In this paper we study a minimal model of QOHM which is universal
\citep{Gelbwaser-Klimovsky2013,Myers2020} in the sense that it can
works either as a heat engine or a refrigerator or, yet, a heater
-- see Fig. \ref{Fig1}, depending on the control parameter. Our
model consists of a two-level system (TLS) driven by an external laser
source and \textcolor{black}{interacting with a cold thermal reservoir
and with a squeezed hot thermal reservoir, which will be assumed as
a free resource \citep{Klaers2017}. The relation between }$\eta$
and $COP$\textcolor{black}{{} (Eq. (\ref{eq:eq1})) will be generalized
to include the squeezing parameter, which will be our parameter of
control in building these types of heat machines. }Using our model,
we are able to study both the efficiency and performance of this TLS
machine at finite-time regime of the isentropic strokes, which contributes
to making them useful from the point of view of applicability.

This paper is organized as follows. In Sec. \ref{sec:II} we present
our model for a universal QOHM, which consists of a TLS as the working
substance under a cold thermal and a squeezed hot thermal reservoir.
In Sec. \ref{sec:III} we present the results of our calculation for
the heats exchanged with the reservoirs and the work done or performed
by the QOHM and the corresponding efficiency $\eta$ and performance
coefficient $COP$. In Sec. \ref{sec:IV} we generalize \textcolor{black}{the
relation between }$\eta$ and $COP$ given by Eq. (\ref{eq:eq1})
to include both the squeezed thermal reservoir and finite-time isentropic
strokes. Finally, in Sec. \ref{sec:V} we present our conclusions.

\section{\label{sec:II}Universal QOHM}

\textcolor{black}{We are going to consider the TLS implementation
of a four-stroke quantum Otto cycle.} The four-stroke to our QOHM
are the following:

\emph{(i) Cooling stroke.} In this first step, the TLS is weakly coupled
to the cold thermal reservoir until thermalized, when it can be described
by the Gibbs state $\rho_{1}^{G}=\text{e}^{-\beta_{c}H_{c}}/\text{Tr}\left(\text{e}^{-\beta_{c}H_{c}}\right)$,
where $H_{c}$ is the Hamiltonian and $\beta_{c}=1/k_{B}T_{c}$, where
$k_{B}$ is the Boltzmann constant and $T_{c}$ is the reservoir temperature.
The TLS Hamiltonian remains unchanged during the thermalization process
and has the form $H_{c}=\frac{1}{2}\hbar\omega_{c}\sigma_{x}$, with
$\hbar$, $\omega_{c}$, and $\sigma_{x}$ being the reduced Planck
constant, the angular frequency, and the x Pauli matrix, respectively.

\emph{(ii) Expansion stroke.} In this stage the TLS evolves unitarily
from the state $\rho_{1}^{G}$ (at time $t=0$) to $\rho_{2}=U\rho_{1}^{G}U^{\dagger}$(at
time $t=\tau$), where $U$ is the unitary operator accounting for
the external driven of the TLS Hamiltonian, which varies from $H_{c}=\frac{1}{2}\hbar\omega_{c}\sigma_{x}$
to $H_{h}=\frac{1}{2}\hbar\omega_{h}\sigma_{y}$, with $\omega_{h}$
being an angular frequency higher then $\omega_{c}$ (corresponding
to the energy gap expansion). For our purpose, it is not necessary
to specify the unitary operator $U$.

(iii) Heating stroke. This is the stage where the TLS is weakly coupled
to the hot squeezed  thermal reservoir until reaching the stead state
$\rho_{3}^{S}=S\rho_{3}^{G}S^{\dagger}$. The reservoir squeezing
changes the thermalized state $\rho_{3}^{G}=\text{e}^{-\beta_{h}H_{h}}/\text{Tr}\left(\text{e}^{-\beta_{h}H_{h}}\right)$
according to operator $S=\left(\mu\left|-_{y}\right\rangle \left\langle +_{y}\right|+\nu\left|+_{y}\right\rangle \left\langle -_{y}\right|\right)/\sqrt{\mu^{2}+\nu^{2}}$,
where $\mu=\cosh r$ and $\nu=\sinh r$ \citep{Srikanth2008}. The
state $\left|-_{y}\right\rangle $ ($\left|+_{y}\right\rangle $)
is the ground (excited) state of the TLS at this stage, and $r$ is
the \emph{squeezing parameter}. As in the cooling stroke, here the
TLS Hamiltonian $H_{h}=\frac{1}{2}\hbar\omega_{h}\sigma_{y}$ also
remains unchanged.

(iv) Compression stroke. This stage is accomplished by reversing the
expansion protocol (ii), such that the TLS Hamiltonian is changed
from $H_{h}=\frac{1}{2}\hbar\omega_{h}\sigma_{y}$ to $H_{c}=\frac{1}{2}\hbar\omega_{c}\sigma_{x}$,
corresponding to the energy gap compression, making the TLS state
to evolve unitarily from to $\rho_{3}^{S}$ to $\rho_{4}=U^{\dagger}\rho_{3}^{S}U$.

The quantities we are interested in is the efficiency $\eta$ to the
heat engine as well as the coefficient of performance $COP$ to the
refrigerator.\textbf{ }The engine efficiency is\textbf{ }given by
$\eta=-W_{net}/Q_{h}$, where $Q_{h}$ is the average heat absorbed
from the hot reservoir and $W_{net}$ is the average net work extracted
from the engine, while the coefficient of performance, on the other
hand, is given by $COP=Q_{c}/W_{net}$, where $Q_{c}$ is the average
heat extracted from the cold thermal reservoir by performing an average
net work $W_{net}$ on the TLS. In order to determine both the efficiency
and the coefficient of performance we resort to the first law of thermodynamics,
together with work and heat definitions, as follows. According to
the first law of thermodynamics, the change in the internal energy
of a given system during a thermodynamic process can be decomposed
into work $W$ and heat $Q$. In quantum thermodynamics the first
law is written as $\Delta E=Q+W$, where $\Delta E$ is the average
change in the system internal energy, which is given by $E=\text{Tr}\left(\rho H\right)$.
Heat and work averages, in turn, are $Q=\int dt\text{Tr}\left[\left(d\rho/dt\right)H\right]$
and $W=\int dt\text{Tr}\left[\rho\left(dH/dt\right)\right]$ \citep{Alicki1979,Kosloff1984},
respectively. Therefore, according to the first law, from the quantum
Otto cycle described in (i)-(iv), we can promptly see that $W=0$
and $\Delta E=Q$ in the heating and cooling strokes, and $Q=0$ and
$\Delta E=W$ in the expansion and compression strokes. This considerably
simplifies the calculations, as compared to other cyclic machines
as for example Carnot or Stirling, where work and heat are simultaneously
exchanged. 

Aiming at possible applications in nuclear magnetic resonance \citep{Peterson2019,Assis2019},
we use here the following parameters in our numerical calculations:
$\omega_{c}=2\pi$ kHz, $\omega_{h}=3.5\omega_{c}$, $\beta_{c}=1/\left(10\ \text{peV}\right)$
, and $\beta_{h}=0.7\beta_{c}$.

\section{\label{sec:III}Results}

With the information provided in (i)-(iv) strokes and definitions
of work and heat as given in Sec. \ref{sec:II}, we can obtain the
average heats exchanged with the cold and hot reservoirs as well as
the average net work: 
\begin{equation}
Q_{c}=-\frac{1}{2}\hbar\omega_{c}\left(\tanh\theta_{c}-\zeta\tanh\theta_{h}\right)-\hbar\xi\zeta\omega_{c}\tanh\theta_{h}\label{Qc}
\end{equation}
\begin{equation}
Q_{h}=\frac{1}{2}\hbar\omega_{h}\left(\tanh\theta_{c}-\zeta\tanh\theta_{h}\right)-\hbar\xi\omega_{h}\tanh\theta_{c},\label{Qh}
\end{equation}
and 
\begin{multline}
W_{net}=-\frac{1}{2}\hbar\left(\omega_{h}-\omega_{c}\right)\left(\tanh\theta_{c}-\zeta\tanh\theta_{h}\right)\\
+\hbar\xi\left(\omega_{h}\tanh\theta_{c}+\zeta\omega_{c}\tanh\theta_{h}\right),\label{Wnet}
\end{multline}
where $\theta_{c\left(h\right)}=\frac{1}{2}\beta_{c\left(h\right)}\hbar\omega_{c\left(h\right)}$,
$\zeta=1/\left(\mu{}^{2}+\nu{}^{2}\right)^{2}$ and $\xi=\left|\langle\pm_{y}\vert U\vert\mp_{x}\rangle\right|^{2}=\left|\langle\pm_{x}\vert U^{\dagger}\vert\mp_{y}\rangle\right|^{2}$.
The parameter $\xi$, which gives the probability of transition between
the two levels of the TLS, is the \emph{adiabaticity parameter} \citep{Assis2019,Peterson2019}.
This parameter allows us to study the QOHM efficiency and performance
coefficient in any time regime. In fact, this so-called adiabaticity
parameter is the transition probability induced by the unitary evolution
$U$, and the faster the unitary process the greater $\xi$. When
$\xi=0$, the process is called quasi-static and occurs at null power.
Finite-time processes, on the other hand, occurs at non-null power,
and for instantaneous process $U=I$, with $I$ being the identity.
As we shall see, the Otto efficiency and performance coefficient occurs
to $\xi=0$, corresponding to a machine operating at null power. As
we are not attaching any secondary system to exchange work with our
TLS machine, either in the case of heat engines or the in the case
of refrigerators, we can think about this simplified model as a proof
or concept \citep{Peterson2019}, allowing us to impose theoretically
maximum constraints on their efficiency or performance coefficient. 

\begin{figure}[h]
\centering{}\includegraphics{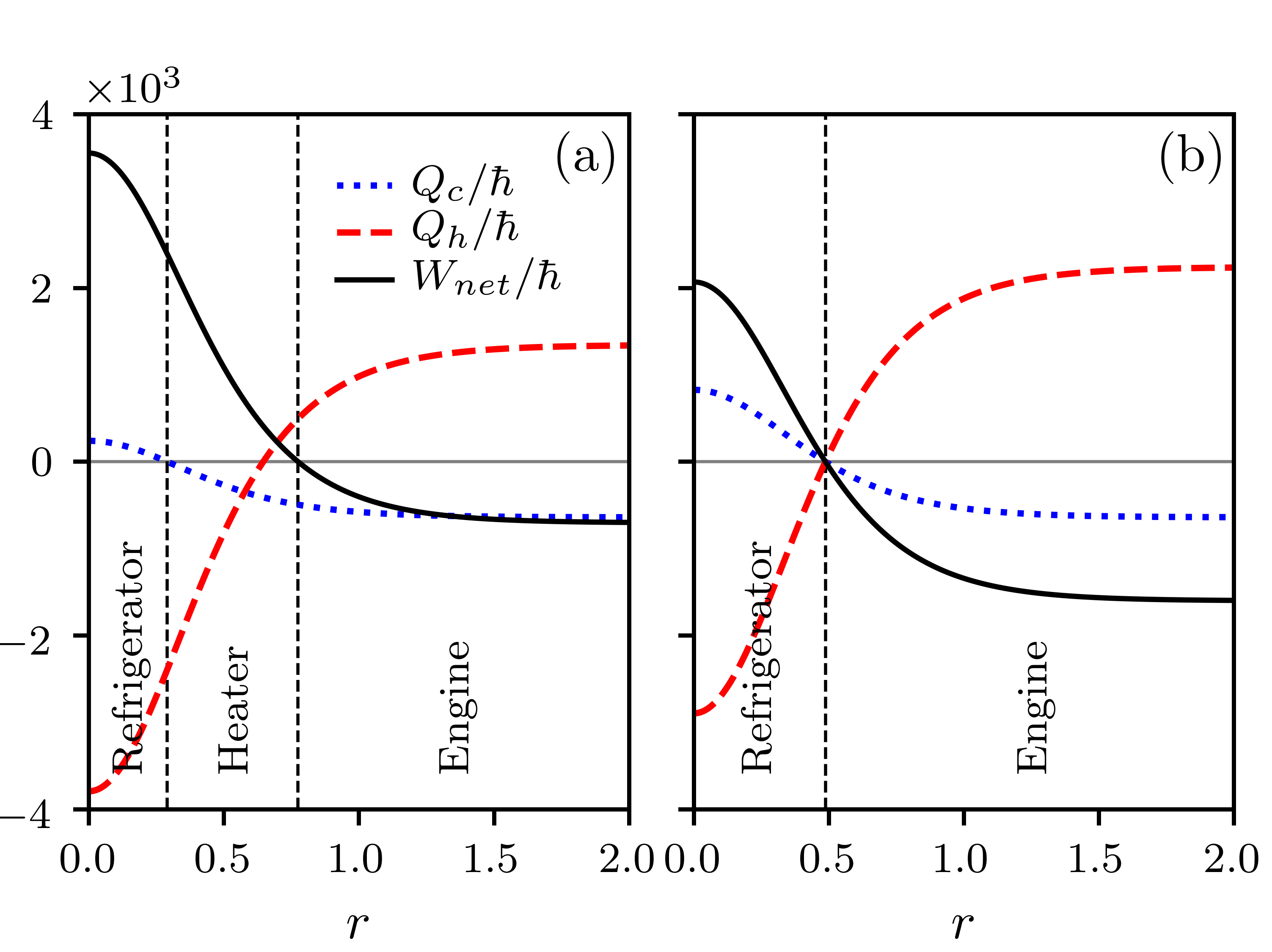}\caption{\label{Fig2} Curves $Q_{c}$ (dotted blue line) $Q_{h}$ (dashed
red line) and $W_{net}$ (solid black line)\emph{ versus} the squeezing
parameter $r$. These curves highlight the universality of our model,
since depending on the value of the control parameter $r$ the four
types of machines shown in Fig. \ref{Fig1} can be engineered. The
parameters used here are (a) $\xi=0.2$, $\omega_{c}=2\pi$ kHz, $\omega_{h}=3.5\omega_{c}$,
$\beta_{c}=1/\left(10\ \text{peV}\right)$ , $\beta_{h}=0.7\beta_{c}$,
and (b) $\xi=0$, $\omega_{c}=2\pi$ kHz, $\omega_{h}=3.5\omega_{c}$,
$\beta_{c}=1/\left(10\ \text{peV}\right)$ , $\beta_{h}=0.7\beta_{c}$.
Dotted black lines delimit the regions of the different types of machine
shown in Fig. \ref{Fig1}.}
\end{figure}

According to our convention, $Q>0$ ($Q<0$) means heat energy flowing
into (out of) the engine, while $W_{net}<0$ ($W_{net}>0$) means
useful energy flowing out of (into) the engine - see Fig. \ref{Fig1}.
In Fig. \ref{Fig2} we show all the three relevant quantities $Q_{c}$
(dotted blue line), $Q_{h}$ (dashed red line) and $W_{net}$ (solid
black line)\emph{ versus} the squeezing parameter $r$ for $\xi=0$
(Fig. \ref{Fig2}(a)) and $\xi=0.2$ (Fig. \ref{Fig2}(b)). From Fig.
\ref{Fig2}(a) the universality of our TLS machine should be apparent.
For example, if we want to build a heat engine, for which $Q_{c}<0$,
$Q_{h}>0$ and $W_{net}<0$, then we should choose $r\apprge0.77$;
if we want to build a refrigerator, for which $Q_{h}<0$, $Q_{c}>0$
and $W_{net}>0$ then our control parameter should be $r\apprle0.29$;
heater machines of types I and II, on the other hand, lies in region
$0.29\apprle r\apprle0.77$, thus corresponding to the four types
of heat machines as shown in Fig. \ref{Fig1}. Also, note from Fig.
\ref{Fig2}(b) that to the quasi-static case $\xi=0$ there are only
two types of heat machines, and depending on the value of $r$, the
machine switches directly from engine to refrigerator and \emph{vice-versa}.

In the following sections we will study the two main types of machines
whose application has been highlighted in the most varied contexts,
which are the engine and the refrigerator, by considering the squeezing
as the control parameter.

\subsection{Two-level heat engine}

Following the definition of efficiency, which is $\eta=-W_{net}/Q_{h}$,
we find to our model:
\begin{equation}
\eta=1-\frac{\omega_{c}}{\omega_{h}}\mathcal{R},\label{eta}
\end{equation}
where 
\begin{equation}
\mathcal{R}=\frac{1+2\xi{\cal F}}{1-2\xi{\cal G}},\label{eq:ratio}
\end{equation}
with 

\begin{equation}
{\cal F}=\frac{\zeta\text{tanh}\theta_{h}}{\text{tanh}\theta_{c}-\zeta\text{tanh}\theta_{h}}\label{F}
\end{equation}
and
\begin{equation}
{\cal G}=\frac{\text{tanh}\theta_{c}}{\text{tanh}\theta_{c}-\zeta\text{tanh}\theta_{h}}.\label{G}
\end{equation}
Note, from Eq. (\ref{eq:ratio}), that $\mathcal{R}=1$ when $\xi=0$,
which corresponds to the quasi-static case, and $\eta=1-\frac{\omega_{c}}{\omega_{h}}\equiv\eta_{Otto}$.

\begin{figure}
\centering{}\includegraphics[width=8.5cm]{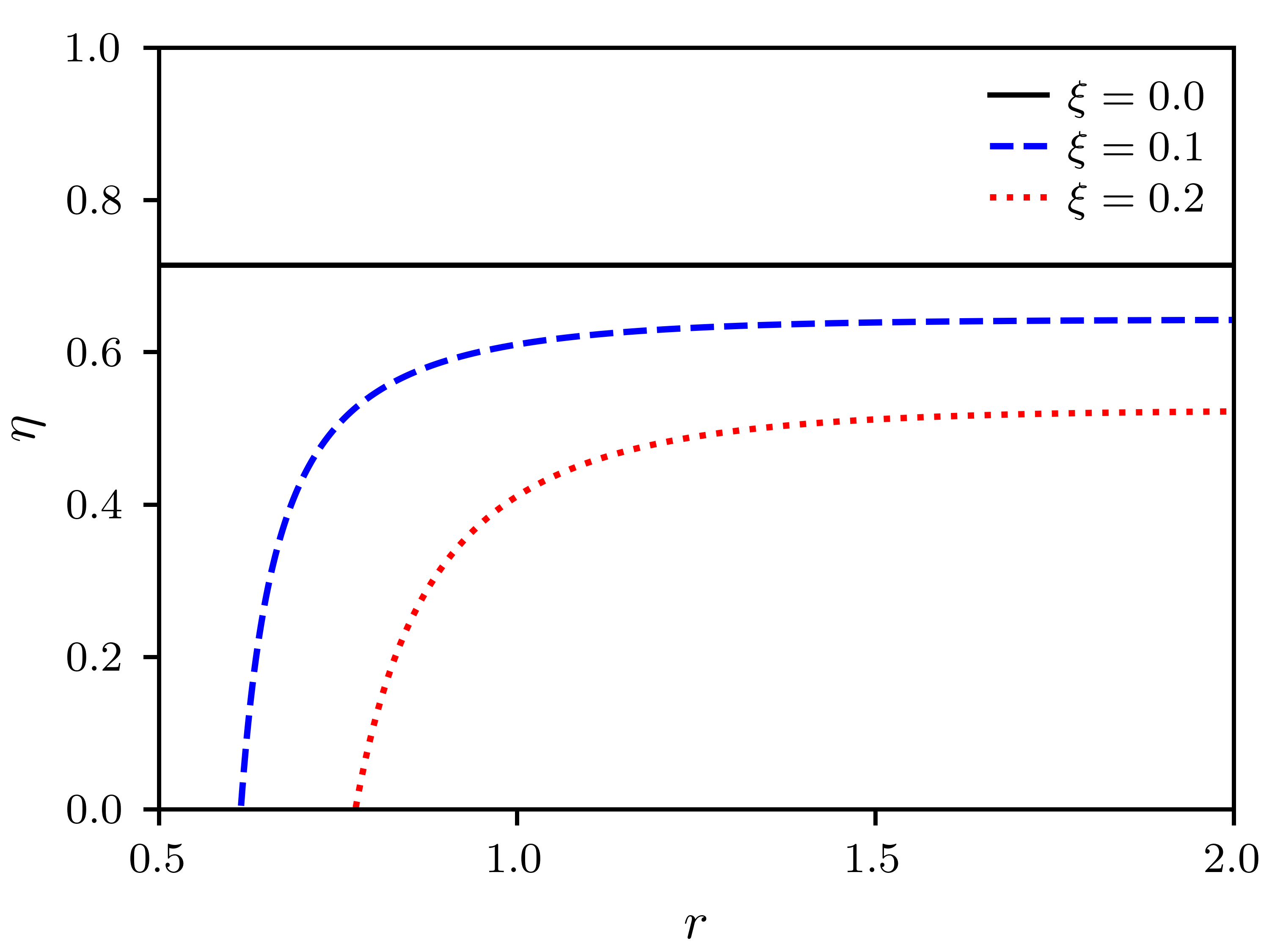}\caption{\label{Fig3} Efficiency \emph{$\eta$ }to the TLS engine\emph{ versus}
the squeezing parameter $r$ considering quasi-static processes $\xi=0$,
resulting in the Otto efficiency $\eta_{Otto}=1-\omega_{c}/\omega_{h}$
(horizontal solid black line), and the finite-time regime $\xi=0.1$
(dashed blue line) and $\xi=0.2$ (dotted red line). We also use the
parameters $\omega_{c}=2\pi$ kHz, $\omega_{h}=3.5\omega_{c}$, $\beta_{c}=1/\left(10\ \text{peV}\right)$,
and $\beta_{h}=0.7\beta_{c}$.}
\end{figure}

In terms of the adiabaticity parameter $\xi$ the condition to extract
work is $W_{net}<0$, such that

\begin{equation}
\xi<\frac{\left(\omega_{h}-\omega_{c}\right)\left(\tanh\theta_{c}-\zeta\tanh\theta_{h}\right)}{2\left(\omega_{h}\tanh\theta_{c}+\zeta\omega_{c}\tanh\theta_{h}\right)},\label{Ad. Parameter}
\end{equation}
which implies $\zeta<\tanh\theta_{c}/\tanh\theta_{h}$ (since $\xi\geq0$).
As we can confirm both analytically and numerically, this condition
results in heat absorption from the squeezed hot thermal reservoir,
$Q_{h}>0$, and heat loss to the cold thermal reservoir, $Q_{c}<0$,
which characterizes the heat engine. In Fig. \ref{Fig3} we show the
efficiency $\eta$ \emph{versus} the squeezing parameter for quasi-static
regime $\xi=0$ (solid black line), which gives the Otto efficiency,
and for finite-time regime $\xi=0.1$ (dashed blue line) and $\xi=0.2$
(dotted red line). Fig. \ref{Fig3} shows that although the engine
efficiency can be enhanced by the squeezed reservoir, the Otto efficiency
is never achieved for processes occurring in finite-time regimes.
However, for a treatment of the engine efficiency when elaborated
optimization procedure is carried out, see Ref. \citep{Assis2020}.

\subsection{Two-level Refrigerator}

According to Eq. (\ref{eq:eq1}), a good engine is a poor refrigerator
and \emph{vice-versa}. This leads us to the conclusion that, as we
have seen in the previous Section, since the squeezing parameter enhances
the engine efficiency, squeezed reservoirs should not enhance the
coefficient of performance. As we shall see, our results confirm that
this is true. The refrigerator $COP$ is defined by the ratio $COP=Q_{c}/W_{net}$,
meaning that the goal is to extract as much heat as possible from
the cold reservoir by doing a minimum of work. From Eqs. (\ref{Qc})
and (\ref{Wnet}) we obtain
\begin{equation}
COP=\frac{\omega_{c}\left(\tanh\theta_{c}-\zeta\tanh\theta_{h}\right)-\xi\zeta\omega_{c}\tanh\theta_{h}}{\left(\omega_{h}-\omega_{c}\right)\left(\tanh\theta_{c}-\zeta\tanh\theta_{h}\right)+\xi\left(\omega_{h}\tanh\theta_{c}+\zeta\omega_{c}\tanh\theta_{h}\right)}
\end{equation}
or, after a little algebra:\textbf{
\begin{equation}
COP=\frac{\mathcal{R}COP_{Otto}}{1+COP_{Otto}\left(1-\mathcal{R}\right)},\label{COP}
\end{equation}
}where $\mathcal{R}$ was defined in Eqs. (\ref{eq:ratio})-(\ref{G})
and 
\begin{equation}
COP_{Otto}=\frac{\omega_{c}}{\omega_{h}-\omega_{c}}
\end{equation}
is the ideal $COP$ obtained from quasi-static processes, i.e., by
letting $\xi=0$ ($\mathcal{R}=1$).

\begin{figure}
\centering{}%
\begin{tabular}{cc}
\includegraphics{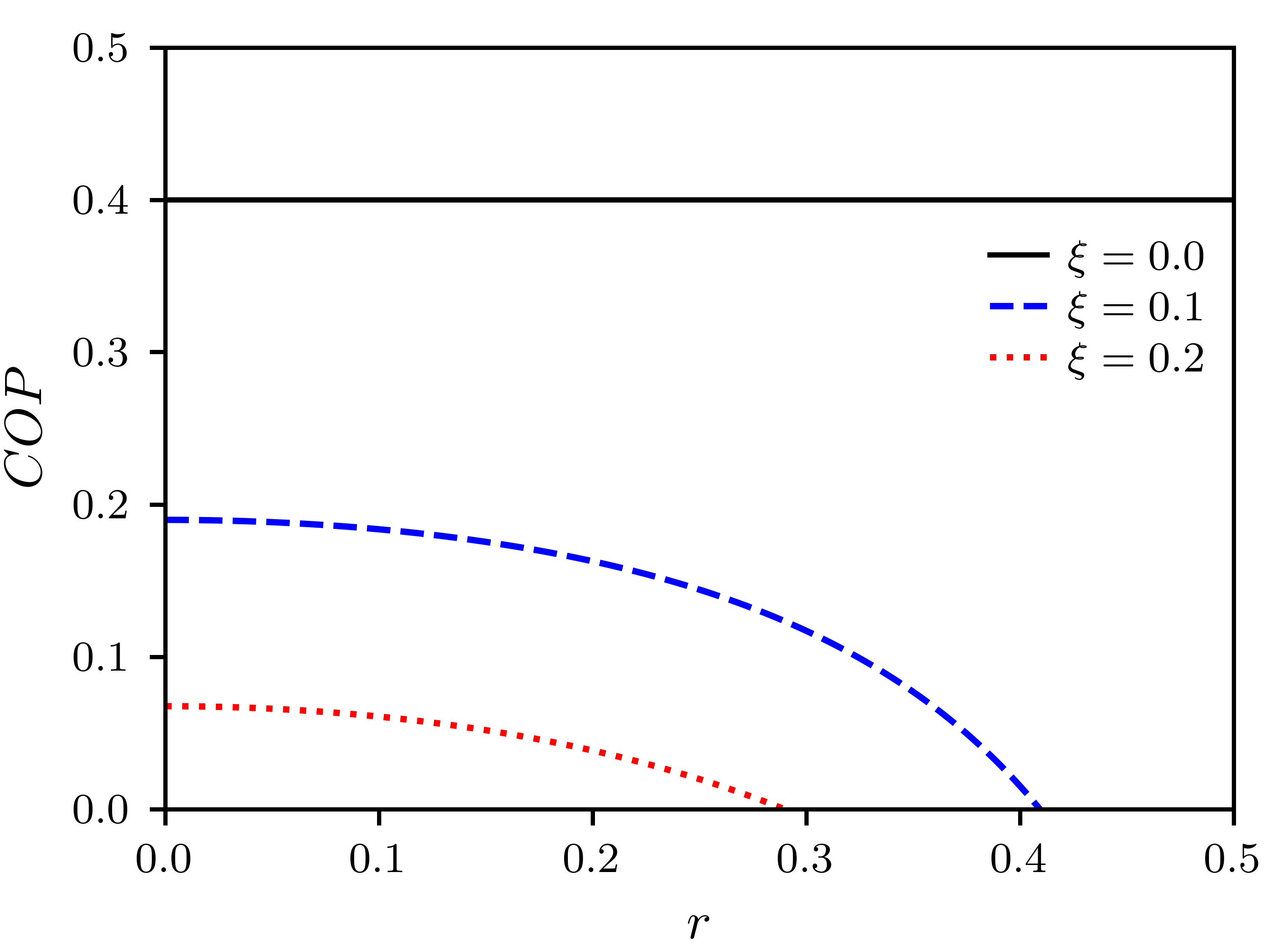} & \tabularnewline
\end{tabular}\caption{\label{Fig4} Performance COP versus the parameter $r$ of the squeezed
reservoir for a (a) quasi-static process $\xi=0$, solid black line,
and finite time processes (b) $\xi=0.1$, dashed blue line, and (c)
$\xi=0.2$, dotted red line. The other parameters used here are $\omega_{c}=2\pi$
kHz, $\omega_{h}=3.5\omega_{c}$, $\beta_{c}=1/\left(10\ \text{peV}\right)$,
and $\beta_{h}=0.7\beta_{c}$.}
\end{figure}

Recalling that Eq. (\ref{COP}) makes sense only for $Q_{c}>0$ and
$W_{net}>0$ , the following constraint must be obeyed: 
\begin{equation}
\xi<\frac{1}{2}\left(1-\frac{\tanh\theta_{c}}{\zeta\tanh\theta_{h}}\right).\label{eq:Constr}
\end{equation}
By imposing the constraint Eq. (\ref{eq:Constr}), we can numerically
verify that the highest value of $COP$ is equal to the ideal Otto
machine $COP_{Otto}=\omega_{c}/\left(\omega_{h}-\omega_{c}\right)$,
which occurs to the quasi-static process $\xi=0$ or $\mathcal{R}=1$.
In Fig. \ref{Fig4} we show the $COP$ for the ideal Otto refrigerator
$\xi=0$, solid black line, as well as for two other finite-time parameters
$\xi=0.1$ (dashed blue line) and $\xi=0.2$ (dotted red line). As
we can see, for $\xi>0$ all curves in Fig. \ref{Fig4} start below
the ideal Otto performance coefficient $COP_{Otto}=0.4$ and decreases
further as the squeezing parameter increases. As expected, the performance
coefficient behavior is contrary to the efficiency behavior shown
in Fig. \ref{Fig3}.

\section{\label{sec:IV}Generalized relation between $\eta$ and $COP$}

In this Section we will generalize Eq. (\ref{eq:eq1}) to take into
account processes occurring at $\xi>0$ and under squeezed reservoirs.
To this end, we eliminate $W_{net}$ from both $\eta=-W_{net}/Q_{h}$
and $COP=Q_{c}/W_{net}$ using Eqs. (\ref{Qc})-(\ref{Wnet}), to
obtain
\begin{equation}
COP=\frac{1}{\eta}\frac{\omega_{c}}{\omega_{h}}\frac{\left[\tanh\left(\theta_{c}\right)-\zeta\tanh\left(\theta_{h}\right)\right]+2\xi\zeta\tanh\left(\theta_{h}\right)}{\left[\tanh\left(\theta_{c}\right)-\zeta\tanh\left(\theta_{h}\right)\right]+2\xi\tanh\left(\theta_{c}\right)},\label{etacop1}
\end{equation}
or, using $\eta_{Otto}=1-\frac{\omega_{c}}{\omega_{h}}$ and $\mathcal{R}$
as defined by Eqs. (\ref{eq:eq1})-(\ref{G}):

\begin{equation}
COP=\left(\frac{1}{\eta}-\frac{\eta_{Otto}}{\eta}\right)\mathcal{R}.\label{etacop2}
\end{equation}
 Note that for $\xi=0$ it is always possible to obtain a relation
between $\eta$ and $COP$, since $\mathcal{R}=1$ and hence $COP=COP_{Otto}$
and $\eta=\eta_{Otto}$:

\begin{equation}
COP_{Otto}=\frac{1}{\eta_{Otto}}-1,
\end{equation}
which is exactly the same as Eq. (\ref{eq:eq1}) and, therefore the
squeezing parameter does not modify the relation between $\eta$ and
$COP$ in the quasi-static limit. However, for $\xi>0$ it should
be noted that, except for a narrow region, see Fig. \ref{Fig2}(a),
in different regions in which the heat machine operates either as
refrigerator or engine, the parameters for which $W_{net}>0$ are
not the same as those for $W_{net}<0$, thus implying that it is not
legitimate to consider $W_{net}^{\left(engime\right)}=-W_{net}^{\left(fridge\right)}$.
As a consequence, unlike what happens in the quasi-static case shown
in Fig. \ref{Fig2}(b), for $\xi>0$ there will not always be a balance
between the work $W_{net}$ that can be extracted from a universal
TLS engine having $\eta$ efficiency, and the COP performance that
would be obtained if that same work $W_{net}$ were supplied to a
fridge machine.

\section{\label{sec:V}Conclusion}

We have proposed a universal quantum Otto heat machine (QOHM) based
on a two-level system as the working substance that operates under
two reservoirs: a cold thermal reservoir and a squeezed hot thermal
reservoir. For universal QOHM we mean the possibility of changing
the parameters of control, such as the squeezing $r$ and the adiabaticity
$\xi$ parameters, to make the machine work either as a thermal engine,
or as a refrigerator, or as a heater. We also showed that the squeezing
parameter, although useful to improve the efficiency of an engine,
always leads to a worsening of the performance coefficient of a refrigerator.
This is in contrast with the result from Ref. \citep{Long2015}, where
the authors considered a harmonic oscillator as the work substance
and the cold reservoir as the squeezed one. Finally, we have demonstrated
that the usual relation between $\eta$ and $COP$, Eq. (\ref{eq:eq1}),
remains unchanged for a heat machine working at null power under two
reservoirs: a cold thermal reservoir and a hot squeezed thermal reservoir. 

\section*{Acknowledgments }

We acknowledge financial support from the Brazilian agency, CAPES
(Financial code 001) CNPq and FAPEG. This work was performed as part
of the Brazilian National Institute of Science and Technology (INCT)
for Quantum Information\textbf{ }Grant No. 465469/2014-0.

\bibliographystyle{apsrev4-1}
\bibliography{IEEEfull,References}

\end{document}